\documentclass[conference]{IEEEtran}
\IEEEoverridecommandlockouts
\usepackage{cite}
\usepackage{amsmath,amssymb,amsfonts}
\usepackage{algorithmic}
\usepackage{graphicx}
\usepackage{textcomp}
\usepackage{amsthm}
\usepackage{xcolor}
\usepackage{bbding}
\usepackage{makecell}
\usepackage{algorithm}  
\usepackage{color}
\usepackage{multirow}
\usepackage{bm}
\usepackage{array}
\usepackage{booktabs}
\usepackage{tabularx,booktabs}
\usepackage{multicol}
\usepackage{cleveref}
\usepackage{balance}
\usepackage{lastpage}
\usepackage{tabulary}
\usepackage{subfigure}
\usepackage{etoolbox}
\usepackage{bbm}
\usepackage{enumitem}
\usepackage{mathrsfs}
\usepackage{multicol}
\usepackage{amsfonts,amssymb}
\usepackage{ulem}
\usepackage{cancel}
\usepackage{setspace}
\usepackage{fancyhdr}
\usepackage{stackengine}
\usepackage{threeparttable}
\usepackage{tcolorbox}

\usepackage{amsthm}
\theoremstyle{definition}

\usepackage{nomencl}
\makenomenclature

\def\BibTeX{{\rm B\kern-.05em{\sc i\kern-.025em b}\kern-.08em
    T\kern-.1667em\lower.7ex\hbox{E}\kern-.125emX}}

\newcommand{\blue}[1]{\textcolor{black}{#1}}

\begin{document}

\bstctlcite{IEEEexample:BSTcontrol}

\title{Generative AI as a Service in 6G Edge-Cloud: Generation Task Offloading by In-context Learning
\thanks{Hao Zhou, Chengming Hu, Dun Yuan, Ye Yuan, and Xue Liu are with the School of Computer Science, McGill University, Montreal, QC H3A 0E9, Canada. (mails:{hao.zhou4, chengming.hu, dun.yuan, ye.yuan3}@mail.mcgill.ca, xueliu@cs.mcgill.ca); Di Wu is with the School of Electrical and Computer Engineering, McGill University, Montreal, QC H3A 0E9, Canada. (email: di.wu5@mcgill.ca); Zhu Han is with the Department of Electrical and Computer Engineering at the University of Houston, Houston, TX 77004 USA. (email: hanzhu22@gmail.com); Jianzhong (Charlie) Zhang is with Samsung Research America, Plano, Texas, TX 75023, USA. (email: jianzhong.z@samsung.com).}
}

\author{\IEEEauthorblockN{ Hao Zhou, Chengming Hu, Dun Yuan, Ye Yuan, Di Wu, \\ Xue Liu, \IEEEmembership{Fellow, IEEE}, Zhu Han, \IEEEmembership{Fellow, IEEE}, and Jianzhong (Charlie) Zhang, \IEEEmembership{Fellow, IEEE}.}}

\maketitle

\thispagestyle{fancy}            
\fancyhead[C] {This paper has been accepted by IEEE Wireless Communications Letters. }

\begin{abstract}
\blue{Generative artificial intelligence (GAI) is a promising technique towards 6G networks, and generative foundation models such as large language models (LLMs) have attracted considerable interest from academia and industry. 
This work considers a novel edge-cloud deployment of foundation models in 6G networks. Specifically, it aims to minimize the service delay of foundation models by radio resource allocation and task offloading, i.e., offloading diverse content generation tasks to proper LLMs at the network edge or cloud.  
In particular, we first introduce the communication system model, i.e., allocating radio resources and calculating link capacity to support generated content transmission, and then we present the LLM inference model to calculate the delay of content generation.
After that, we propose a novel in-context learning method to optimize the task offloading decisions. It utilizes LLM's inference capabilities, and avoids the difficulty of dedicated model training or fine-tuning as in conventional machine learning algorithms.
Finally, the simulations demonstrate that the proposed edge-cloud deployment and in-context learning method can achieve satisfactory generation service quality without dedicated model training.}
\end{abstract}


\section{Introduction}

Generative AI (GAI) has received considerable attention recently, which is capable of analyzing complex data distributions and generating similar new content.
\blue{Due to its promising features, existing studies have started exploring GAI-enabled 6G networks, e.g., GAI for semantic communication \cite{qiao2024latency} and air-to-ground channel modelling \cite{giuliani2024spatially}}.
\blue{As a sub-field of GAI, generative foundation models, especially large language models (LLMs), have attracted interest from both academia and industry. For instance, foundation models have been used for network intrusion detection in \cite{fu2024iov}.
Meanwhile, telecom companies have started applying foundation models, e.g., Apple will bring ChatGPT to iPhones with OpenAI, and Qualcomm has developed a mobile platform to support LLMs.}

The above progress of academia and telecom industry has demonstrated the great potential of GAI foundation models such as LLMs in 6G networks \cite{qiao2024latency,giuliani2024spatially,fu2024iov,liu2024llm,javaid2024leveraging}.
However, despite the advancement, some fundamental and crucial problems are still not investigated, e.g., practical deployment of GAI foundation models within 6G network architecture, and evaluating the service delay of these generation services in wireless environments. 
In particular, large GAI models such as GPT-4 usually have billions of parameters, and practical deployment of these large GAI models is critical to support various applications in wireless networks, i.e., network intrusion detection, generation, and management tasks in \cite{fu2024iov,liu2024llm,javaid2024leveraging}. 
\blue{On the other hand, mobile users may send various service requests over wireless channels with diverse preferences, e.g., question-answering tasks need higher accuracy and chatting tasks expect lower delay. 
Therefore, using one GAI model, i.e., a single LLM, to service the requests of all mobile users is impractical, leading to lower service quality and efficiency.}

To this end, we proposed a novel edge-cloud collaboration deployment strategy. {We consider small-scale LLMs such as Llama3-8B to be deployed at network edge servers of base stations (BSs), aiming to process tasks efficiently with lower delay, e.g., chatting, information extraction, and content summarization.} 
By contrast, large-scale LLMs, e.g., Llama3-70B and GPT-4, are deployed in the central cloud with abundant computational resources. {These large-scale LLMs can generate high-quality content for quality-preferred tasks, such as scientific knowledge, math, and coding-related tasks.}

Such an edge-cloud collaboration enables flexible content generation in wireless networks, but it also involves task offloading decisions, i.e., generating content at network edge or offloading tasks to central cloud. 
Inspired by the recent progress of LLM-based optimization \cite{zhou2024large}, this work further explored in-context learning-based decision-making. Specifically, it uses LLMs to learn from formatted natural language demonstrations and improve the performance on target tasks\cite{dong2022survey}. 
Compared with existing machine learning (ML) methods, in-context learning has several advantages: a) Avoiding the complexity of model training and fine-tuning, a well-known bottleneck of conventional ML techniques; b) Following human language instructions to formulate and solve problems, which is far beyond the capabilities of other ML algorithms.
{In-context learning has been explored to address detection, optimization, prediction tasks, etc \cite{zhou2024large}.}

The core contributions of this work are two-fold:\\
1) Firstly, to the best of our knowledge, this work is the first to model the service delay of foundation GAI models in wireless networks, including the communication models for content transmission, and LLM inference models for content generation. It provides a specified metric to evaluate the delay experienced by mobile users.
2) Secondly, we propose a novel in-context learning method for generation task offloading. It avoids the complexity of dedicated model training and fine-tuning as existing ML techniques, using natural language for network management. 
Finally, the simulations demonstrate that the proposed edge-cloud deployment strategy and in-context learning-based method can achieve satisfactory generation service quality for mobile users.

\section{System Model}

\begin{figure}[t]
\centering
\setlength{\abovecaptionskip}{-2pt} 
\includegraphics[width=0.75\linewidth]{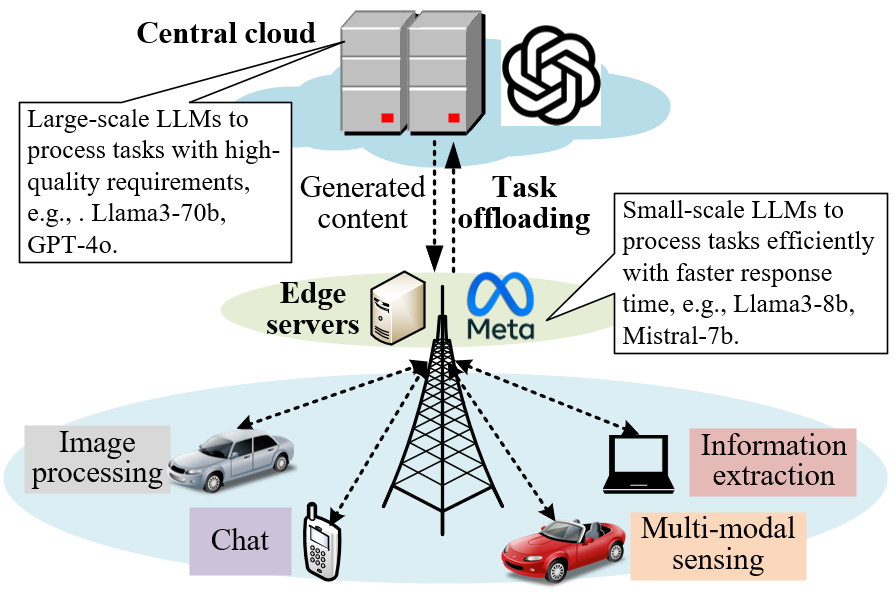}
\caption{\blue{GAI model deployment and services in wireless networks.}}
\vspace{-15pt}
\label{fig-system}
\end{figure}

{Fig. \ref{fig-system} presents the proposed system model, in which mobile users can require various LLM services over wireless networks, e.g., chatting, information extraction and summarization, image generation, etc. Note that these tasks may have different requirements for generated content qualities.} 
Small-scale LLMs are deployed at the network edge to process tasks efficiently with lower delay. By contrast, large-scale LLMs are deployed in the central cloud to handle tasks with high-quality requirements. 
Therefore, the edge servers must make offloading decisions properly, i.e., processing generation tasks locally or offloading them to the central cloud.

\subsection{\blue{Communication Model for Content Transmission}}

\blue{The communication system model considers a downlink transmission for the generated content, involving the wireless transmission delay from BS to end-users, and possible backhaul delay from the network edge to the central cloud as shown in Fig. \ref{fig-system}.}
\blue{The transmission delay $t^{trans}_{k,i}$ for downloading the generated content $i$ for user $k$ is}
\begin{equation}\label{eq3}
t^{tran}_{k,i}=\frac{n^{token}_{k,i}s^{token}}{C_{j,k}}+\alpha_{k,i} t^{back},   
\end{equation}
where $n^{token}_{k,i}$ is the LLM output token numbers for input prompt $i$, $C_{j,k}$ is the link capacity for download transmission between user $k$ and BS $j$, $s^{token}$ is the byte size per token\cite{token_size}. $\frac{n^{token}_{k,i}s^{token}}{C_{j,k}}$ represents the transmission delay from BSs to users. $t_{back}$ is the backhaul delay caused by task offloading. $\alpha_{k,i}$ is the task offloading decision: $\alpha_{k,i}=1$ means offloading tasks to central cloud, while $\alpha_{k,i}=0$ indicates network edge implementation. We assume a fixed $t_{back}$ in (\ref{eq3}), which is a setting used in many task offloading-related studies\cite{el2019macro}.   
The link capacities $C_{j,k}$ can be calculated by
\begin{equation}
\label{eq4}
\resizebox{0.91\hsize}{!}{$
C_{j,k}=\sum\limits_{q\in{\mathcal{Q}_{j}}}b_{q}log(1+ \frac{p_{j,q}g_{j,q,k}z_{j,q,k}}{\sum\limits_{j'\in J_{-j}}{p_{j',q'}g_{j',q',k'}z_{j',q',k'}}+b_{q}N_{0}}),$}
\end{equation}
where $\mathcal{Q}_{j}$ is the resource blocks (RBs) set of the $j^{th}$ BS, $b_{q}$ is the bandwidth of the $q^{th}$ RB, $p_{j,q}$ is the transmission power of the $q^{th}$ RB,  $g_{j,q,k}$ is the channel gain between BS and user, $z_{j,q,k}$ is a binary indicator to represent whether the $q^{th}$ RB is allocated to user $k$, and $N_{0}$ is the noise power density. $J_{-j}$ is the set of BSs except $j^{th}$ BS, and $\sum_{j'\in J_{-j}}{p_{j',q'}g_{j',q',k'}z_{j',q',k'}}$ is inter-cell interference. We assume orthogonal frequency-division multiplexing is deployed to avoid intra-cell interference.

\begin{figure}[t]
\centering
\setlength{\abovecaptionskip}{-5pt} 
\includegraphics[width=0.9\linewidth]{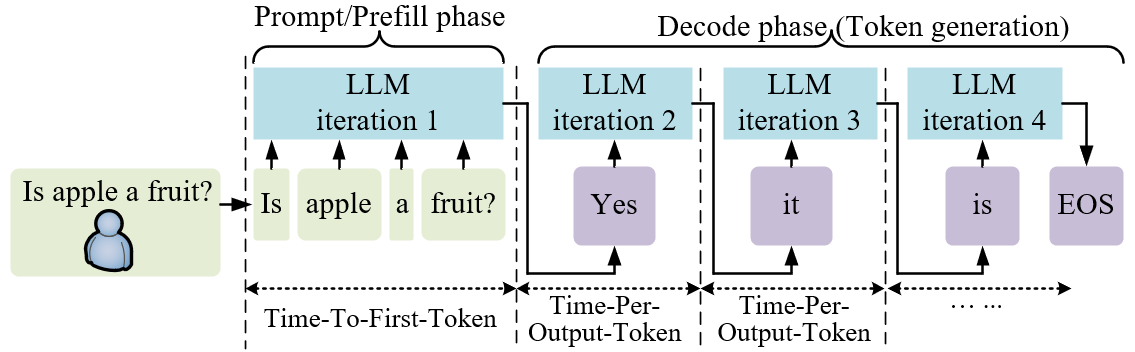}
\caption{\blue{LLM inference process illustration. (EOS: end-of-sequence)}.}
\label{fig-inference}
\vspace{-15pt}
\end{figure}

\subsection{\blue{LLM Inference Model for Content Generation}}

\blue{As shown in Fig. \ref{fig-inference}, the LLM inference process mainly consists of prefill phase and decode phase \cite{patel2023splitwise}.
The user question is split into smaller tokens, then the LLM processes the input tokens as a next-token predictor by autoregressive decoding.}
Fig. \ref{fig-inference} demonstrates that the LLM inference time can be generally divided into two parts: Time-To-First-Token (TTFT) refers to the time to generate the first token, and Time-Per-Output-Token (TPOT) indicates the time to generate each following token \cite{patel2023splitwise}. 
Therefore, the total generation time for a prompt $i$ from user $k$ is
\begin{equation}\label{eq1}
t^{gen}_{k,i}=t^{TTFT}+n^{token}_{k,i}t^{TPOT},   
\end{equation}
where $t^{gen}_{k,i}$ is the total generation time, $t^{TTFT}$ is TTFT time, $n^{token}_{k,i}$ is the number of tokens generated for prompt $i$, and $t^{TPOT}$ is the TPOT time. 
LLM is a complicated system with a huge number of parameters, and $t^{TTFT}$ and $t^{TPOT}$ are affected by many factors, e.g., model architecture, hardware constraints, and task types. Therefore, it is extremely difficult to calculate the exact generation time for each task.
However, (\ref{eq1}) provides a practical approach to quantify LLM generation time since the TTFT and TPOT values of many LLMs can be easily tested and obtained for evaluation purposes \cite{llm_website}. 


\begin{figure*}[t]
\centering
\setlength{\abovecaptionskip}{-5pt} 
\includegraphics[width=0.9\linewidth]{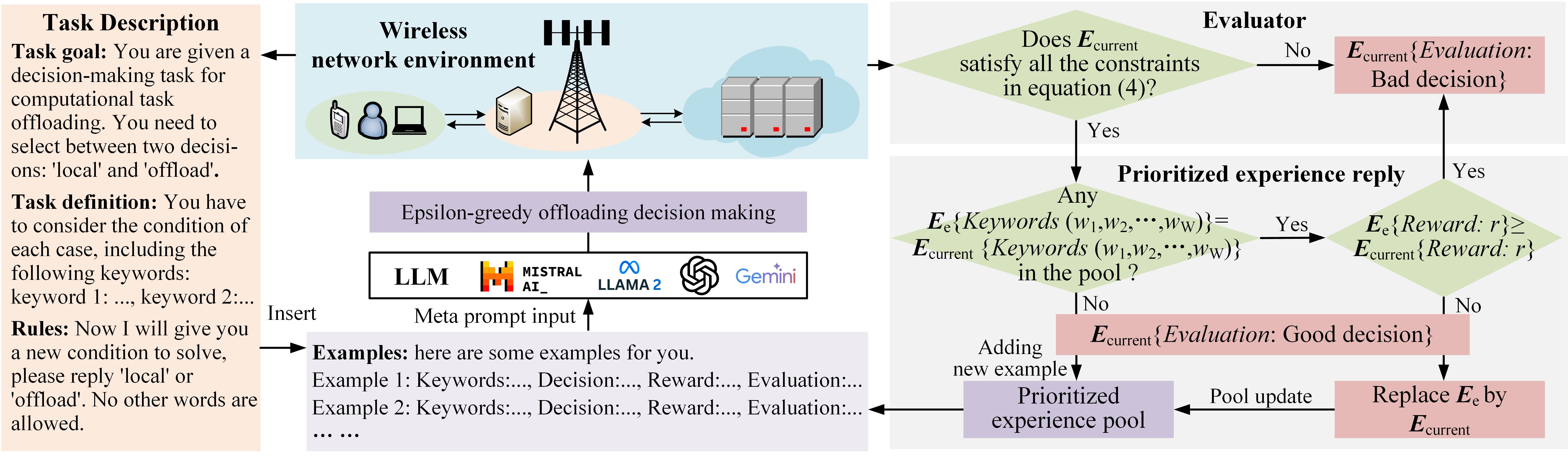}
\caption{LLM-enabled in-context learning for task offloading.}
\label{fig-optimization}
\vspace{-15pt}
\end{figure*}

\subsection{Problem Formulation}
\blue{This work aims to minimize the total generation and transmission delay of all $K$ users from BS $j$ in wireless networks, and meanwhile satisfy the quality requirements of the generated content}. The problem formulation is defined as
\begin{subequations}\label{e5:main}
\begin{align}
\min\limits_{\alpha_{k,i}, \atop z_{j,q,k}} & \enspace \sum_{k=1}^{K} \sum_{i=1}^{I_{k}}  \Big( t^{tran}_{k,i}+ \alpha_{k,i}t^{cloud}_{k,i} +(1-\alpha_{k,i})t^{edge}_{k,i} \Big)   & \tag{\ref{e5:main}} \\
\text{s.t.}  \quad & (\ref{eq3}) \text{ and } (\ref{eq4}),  & \label{e5:c}\\ 
& z_{j,q,k}, \alpha_{k,i} \in \{0,1\}, & \label{e5:a}\\
& t^{edge}_{k,i}=t^{TTFT,edge}+n^{token}_{k,i}t^{TPOT,edge}, & \label{e5:e}\\ 
& t^{cloud}_{k,i}=t^{TTFT,cloud}+n^{token}_{k,i}t^{TPOT,cloud},  & \label{e5:d}\\
& \tau_{k,i} \leq \alpha_{k,i} \tau^{cloud} +(1-\alpha_{k,i}) \tau^{edge},  & \label{e5:b}
\end{align}
\end{subequations}
where $\alpha_{k,i}$ is the task offloading decision, $z_{j,q,k}$ is the RB allocation decision defined in (\ref{eq4}), and $I_{k}$ is the total number of service requests from user $k$. 
We assume a small-scale LLM is deployed at the network edge as in constraint (\ref{e5:e}), in which $t^{edge}_{k, i}$, $t^{TTFT, edge}$, and $t^{TPOT, edge}$ represent total generation time, TTFT, and TPOT values. Similarly, $t^{cloud}_{k,i}$, $t^{TTFT, cloud}$, and $t^{TPOT, cloud}$ are defined for the cloud LLM in (\ref{e5:d}).     
\blue{In addition, the quality of the generated content is also a crucial metric to evaluate generation services.}
(\ref{e5:b}) is the generation quality constraint, indicating that the quality index of selected edge LLM $\tau^{edge}$ or cloud LLM $\tau^{cloud}$ should be higher than the user requirement $\tau_{k,i}$\footnote{The generation quality index can be obtained by testing LLMs on task datasets, e.g., Chatbot Arena, Multi-task Language Understanding, etc \cite{llm_website}.}.

\blue{Finally, (\ref{e5:main}) includes both radio resource allocation and task offloading decisions, in which $z_{j,q,k}$ for radio resource allocation and $\alpha_{k,i}$ for task offloading. For radio resource allocation,} we apply a classic proportional fairness algorithm because: 1) proportional fairness is a practical method that has been widely used in many existing studies; 2) this work aims to understand generative foundation model service properties, and it is reasonable to apply a well-known resource allocation method to better focus on foundation models. For task offloading, we propose an in-context learning-based task offloading approach in the following Section \ref{sec-conext}.

\section{In-context Learning for Task Offloading}
\label{sec-conext}


In-context learning refers to the process of learning from formatted natural language-based task descriptions and examples, aiming to improve the performance of target tasks~\cite{dong2022survey}.    
Considering a query input $x$ and possible candidate answers $\mathcal{Y}=\{y_1, y_2,...,y_{|\mathcal{Y}|}\}$, a set of examples are provided as $\mathcal{E}=\{E_1, E_2,...,E_{|\mathcal{E}|}\}$, in which each $E_e \in \mathcal{E}$ consists of input-output pairs as $E_e=(x_e,y_e)$.
The probability of generating a specific output $y^{*}$ is
\begin{equation}\label{eq7}
Pr(y^{*}|x) \triangleq f_{LLM}(x, y^{*}, \{E_1, E_2,...,E_{|\mathcal{E}|}\}, D),   
\end{equation}
where $f_{LLM}(\cdot)$ is a scoring function and $D$ is the task description.
Then the final output answer $\hat{y}$ is the candidate answer with the highest probability
\begin{equation}\label{eq8}
\hat{y} = \arg \max\limits_{y \in \mathcal{Y}}(Pr(y|x)).   
\end{equation}
The above (\ref{eq7}) and (\ref{eq8}) prove that the output $\hat{y}$ depends on the input $x$, task description $D$, and example set $\mathcal{E}$. 
In this work, LLM outputs $\hat{y}$ refers to the decision between local implementation and offloading, the input $x$ involves service types and the estimated output token size, $D$ is the offloading task description, and $\mathcal{E}$ is a set of previous examples, which will be introduced in following subsections. 

\subsection{Prompting System Design}
The overall organization of the proposed LLM-enabled in-context learning is shown in Fig. \ref{fig-optimization} with the following steps:

\textbf{Step 1: Task description.} Based on network environments, the task description $D$ in (\ref{eq7}) is first defined by 
\begin{equation} \label{eq9}
D=\{\textit{Task goal}, \textit{Task definition}, \textit{Rules}\},
\end{equation}
in which the “\textit{Task goal}” specifies the target problem with two decision variables “\textit{local}” and “\textit{offload}”. The “\textit{Task definition}” indicates the status variables affecting offloading decisions. 
Here we consider “\textit{Service types}” and “\textit{Estimated output token size}”, which also means the input $x$ in (\ref{eq7}) is 
\begin{equation} \label{eq10}
x=\{\textit{Service types}, \textit{Estimated output token size}\}.
\end{equation}
Additionally, extra “\textit{Rules}” are applied to LLMs, e.g., replying “\textit{local}” or “\textit{offload}” only to improve the output accuracy.

\textbf{Step 2: Example design.}
The task description $D$ will be combined with the example set  $\mathcal{E}=\{E_1, E_2,..., E_{|\mathcal{E}|}\}$ as a meta prompt input to the LLM, producing task offloading decisions $\alpha$. The example $E_e \in \mathcal{E}$ is defined by
\begin{equation}\label{eq11}
\resizebox{0.91\hsize}{!}{$
\begin{aligned}
E_{e}\{\textit{Keywords}&:(w_1,w_2,...,w_\mathcal{W}), \textit{Decision} \text{: Local/Offload}, \\ 
\textit{Reward}:r,& \textit{ Evaluation: } \text{Good/Bad decision.}\} 
\end{aligned}$} 
\end{equation}
in which the “\textit{Keywords}” refer to the values of “\textit{Service types}” and “\textit{Estimated output token size}” defined in (\ref{eq10}). 
Inspired by reinforcement learning, a reward metric is defined to evaluate the system performance by jointly considering service delay and quality requirements.  
\begin{equation}\label{eq12}
r=  T^{Target}- t^{total}- r^{penalty} 
\end{equation}
where $T^{Target}$ is the target delay, and $t^{total}$ is the objective function defined in (\ref{e5:main}) and  $t^{total}=t^{tran}_{k,i}- \alpha_{k,i}t^{cloud}_{k,i} -(1-\alpha_{k,i})t^{edge}_{k,i}$. 
%
{Here $r^{penalty}$ is a penalty item, aiming to balance the service delay and generated content quality.  
$r^{penalty}$ is a preset positive value if the constraint in (\ref{e5:main}) is violated. A high $r^{penalty}$ value means strict requirements on generated content quality, while a lower $r^{penalty}$ value allows lower generation quality to achieve lower service delay.    
Otherwise, if all constraints are satisfied in (\ref{e5:main}), $r^{penalty}=0$. Such a definition is also a widely used approach in reinforcement learning studies to handle optimization constraints. }
Therefore, (\ref{eq12}) provides a comprehensive metric to minimize the total delay under the constraints.

\textbf{Step 3: Experience evaluation and replay.}
Given the offloading decision, the current network operation results become a new experience example $E_{current}$, which will be sent to the evaluator module as in Fig. \ref{fig-optimization}. 
Specifically, $E_{current}$ is considered a bad decision if the constraints in (\ref{e5:main}) are not satisfied; otherwise, $E_{current}$ is sent to the prioritized experience replay module for further evaluation, which will be introduced in the following Subsection \ref{sec-pool}.

\textbf{Step 4: Meta prompt updating}.
The experience pool in Step 3 will serve as a new example set $\mathcal{E}_{next}$, and the LLM will generate the next decision using the same task description $D$ and the updated example set $\mathcal{E}_{next}$. 
{Steps 3 and 4 will be repeated until algorithm performance converges, which will be analyzed in the following Fig. \ref{fig-conver}.} 

\subsection{Prioritized Experience Replay and Exploration Strategies}
\label{sec-pool}
The above Steps 3 and 4 show that properly updating the experience pool is crucial, and this subsection will present two key techniques, namely prioritized experience replay and epsilon-greedy exploration, aiming to identify the most useful examples for experience replay. 
In particular, Fig. \ref{fig-optimization} shows that the prioritized experience replay includes two rules: 

1) If $E_{current}$ is a new example, which means that 
$E_{current}\{\textit{Keywords:}(w_1,w_2,...,w_\mathcal{W})\}\neq E_{e}\{\textit{Keywords:}(w_1,$ $w_2,...,w_\mathcal{W})\}$ for $\forall$  
$E_{e} \in \mathcal{E}$, then $E_{current}$ is always considered as a “\textit{Good decision}” since there is no existing example of this condition in the current experience pool $\mathcal{E}$.

2) If $\exists E_{e} \in \mathcal{E}$ that has the same keyword values as $E_{current}$ with $E_{e}\{\textit{Keywords:}(w_1,w_2,...,w_\mathcal{W})\}= E_{current}$ $\{\textit{Keywords:}(w_1,$ $w_2,...,w_\mathcal{W})\}$, then we will compare their reward values. 
If $E_{e}$ has a higher reward than $E_{current}$, then $E_{current}$ is considered a “\textit{Bad decision}” and the experience pool $\mathcal{E}$ remains unchanged. 
Otherwise, if $E_{current}$ has a higher reward, then $E_{current}$ becomes a better example as a “\textit{Good decision}". $E_{e}$ will be replaced by $E_{current}$ in the experience pool, and a new example set $\mathcal{E}_{next}$ is generated.

\begin{figure}[t]
\centering
\setlength{\abovecaptionskip}{-5pt} 
\includegraphics[width=1\linewidth]{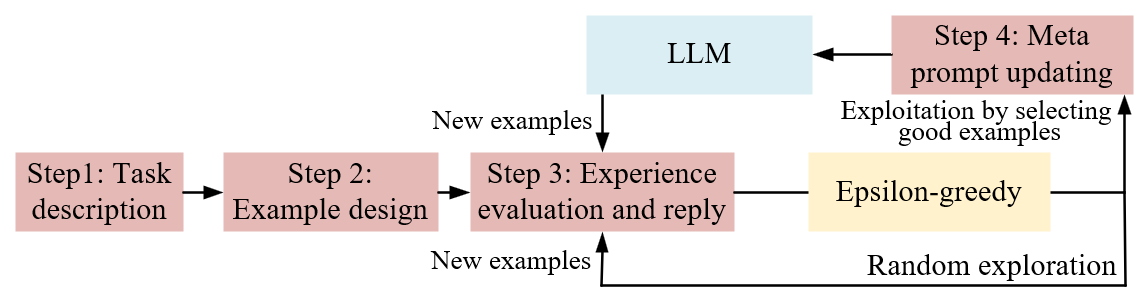}
\caption{{The overall optimization procedure of the proposed technique}}
\label{fig-conver}
\vspace{-15pt}
\end{figure}

{Fig. \ref{fig-conver} summarizes the overall procedure of the proposed techniques. We apply the well-known epsilon-greedy policy to balance exploration and exploitation, which is a fundamental problem in many optimization tasks. Specifically, actions are randomly selected with probability $\epsilon$; otherwise, LLM will make decisions. 
Combining the epsilon-greedy policy with the proposed experience replay technique can send the best examples found to LLMs, and also explore new examples to improve the experience pool.}
{Therefore, with plenty of explorations, the examples in the experience pool will be constantly improved, and the LLM output will converge if no better examples can be found\footnote{{Similar assumptions can be found in the convergence analyses of classic tabular-based Q-learning, e.g., “Q-learning is guaranteed to converge when visiting each state-action tuple infinitely times”.}}. }


Finally, we present 3 baseline algorithms. Baseline 1: Latest experience-based in-context learning, using the latest experience as examples in the prompt. Baseline 2: In-context learning without exploration, and all decisions are made by LLMs. 
Baseline 3: We consider deep reinforcement learning (DRL) as an optimal baseline since DRL techniques have been very widely applied to solve various network optimization problems, in which the state is defined by (\ref{eq10}), the action is offloading decision, and the reward is shown as (\ref{eq12}).

\section{Performance Evaluation}

\begin{figure*}[!t]
\centering
\subfigure[{System reward comparison of different tasks using Llama3-8B and GPT-4o.}]{
\includegraphics[width=5.2cm,height=3.7cm]{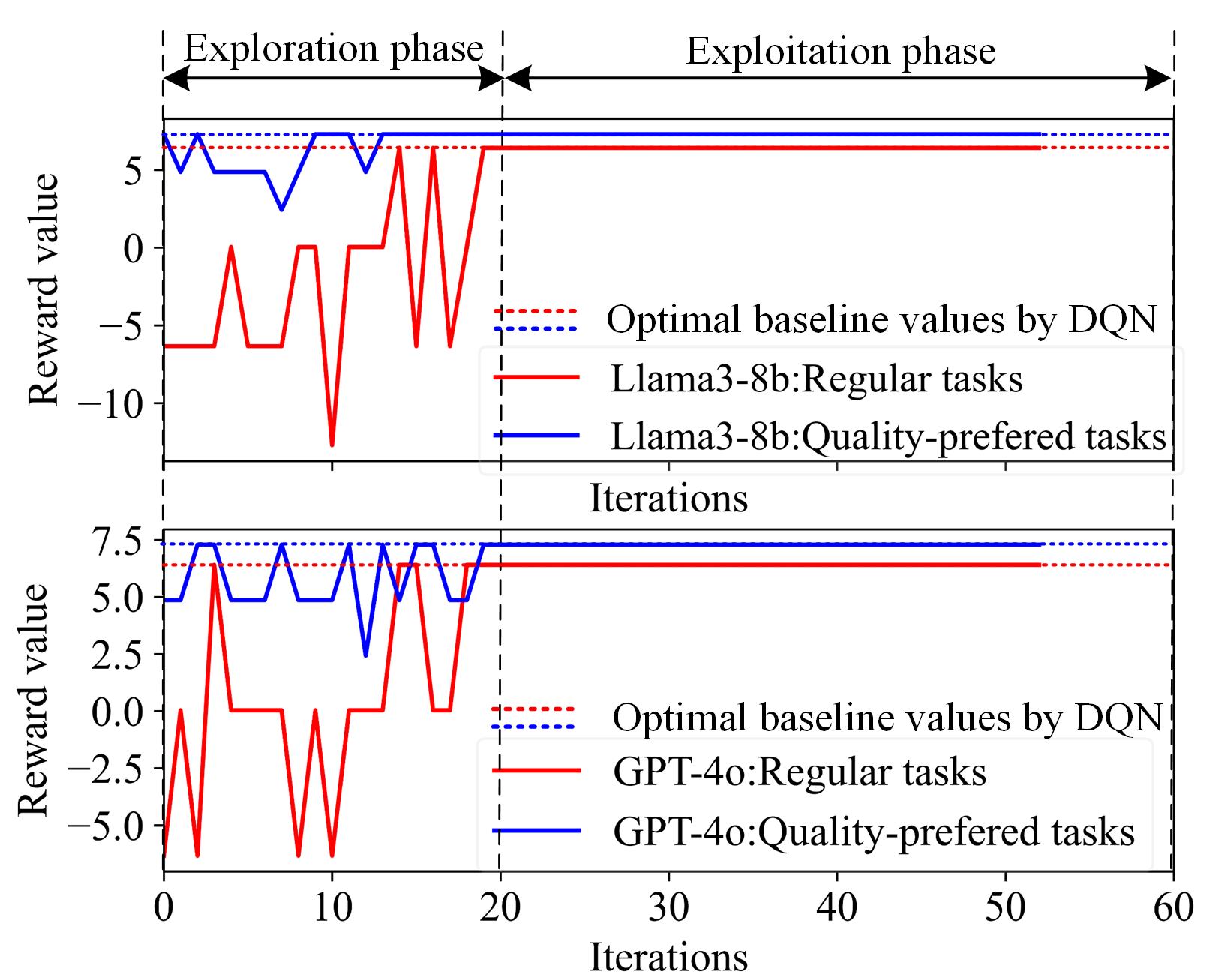} 
\label{f-r1}
}
\subfigure[Service success rate of quality-preferred tasks using using Llama3-8B and GPT-4o. ]{
\includegraphics[width=5.2cm,height=3.7cm]{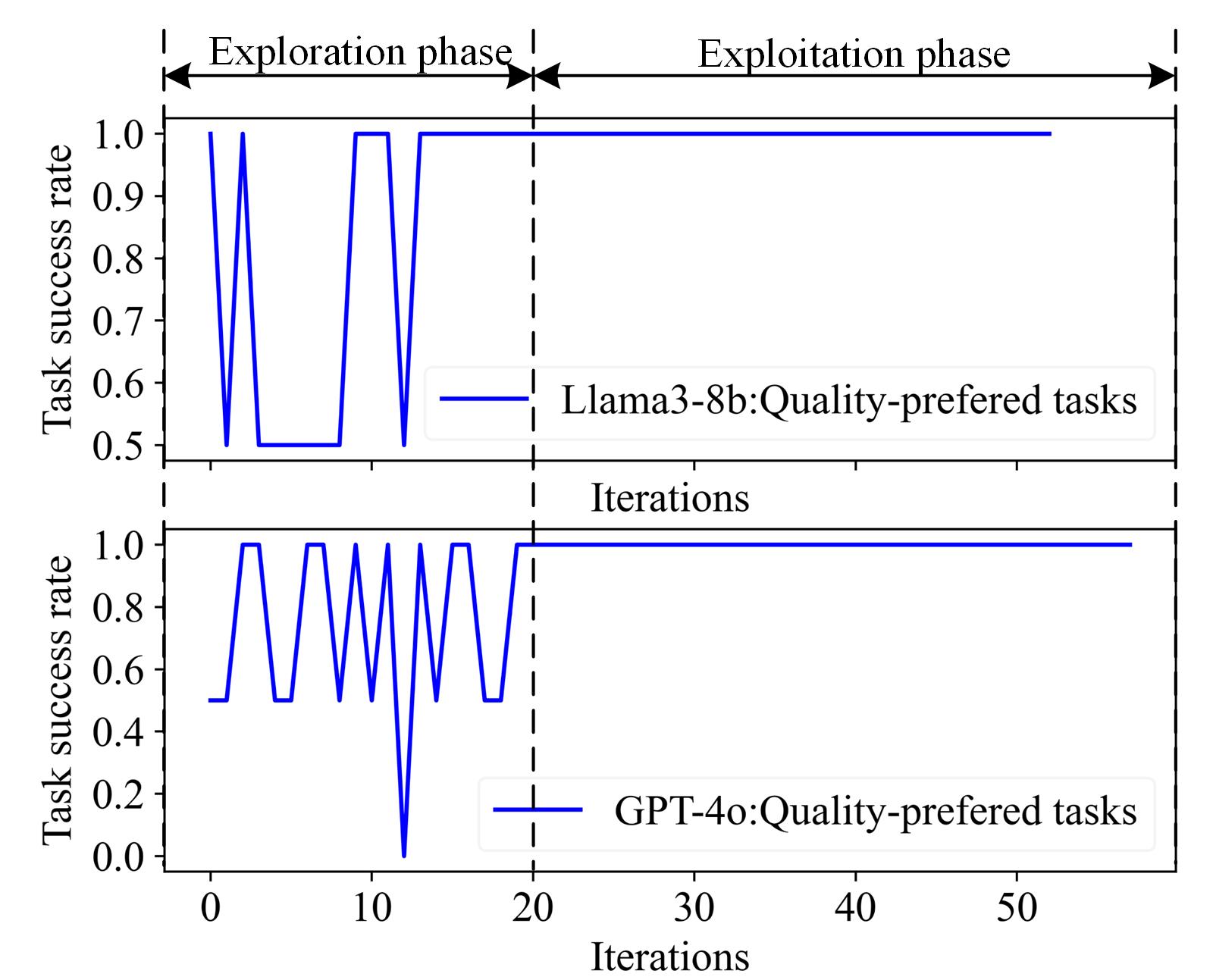} \label{f-r2}
}
\subfigure[{Comparisons under different experience replay and exploration methods using Llama3-8B.}]{
\includegraphics[width=5.2cm,height=3.7cm]{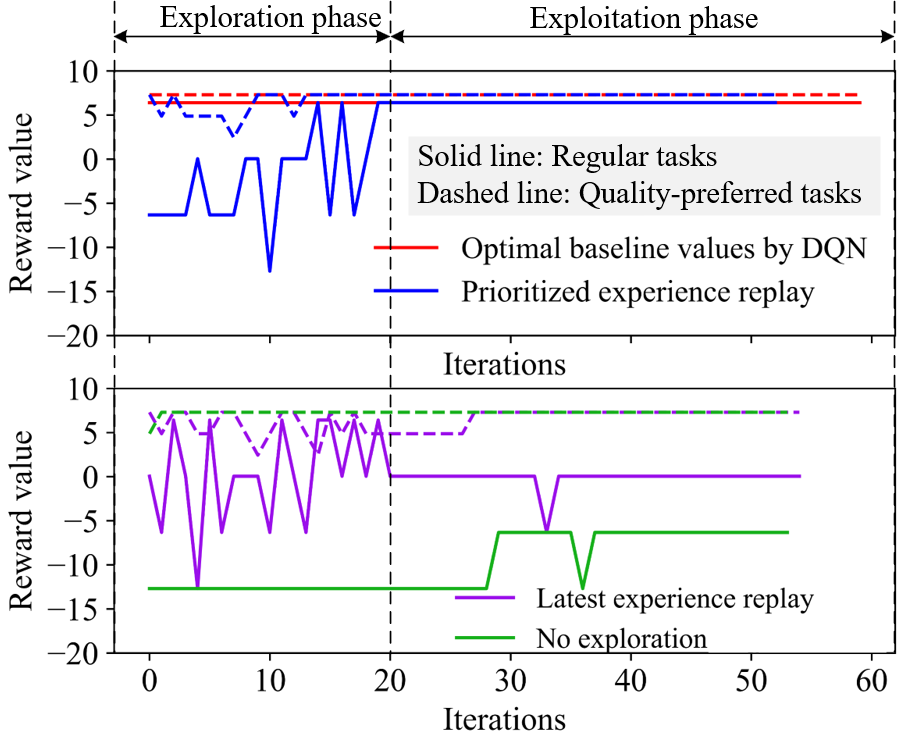} \label{f-r3}
}
\quad
\subfigure[{Average service delay with different user numbers using different LLM combinations.}]{
\includegraphics[width=5.2cm,height=3.7cm]{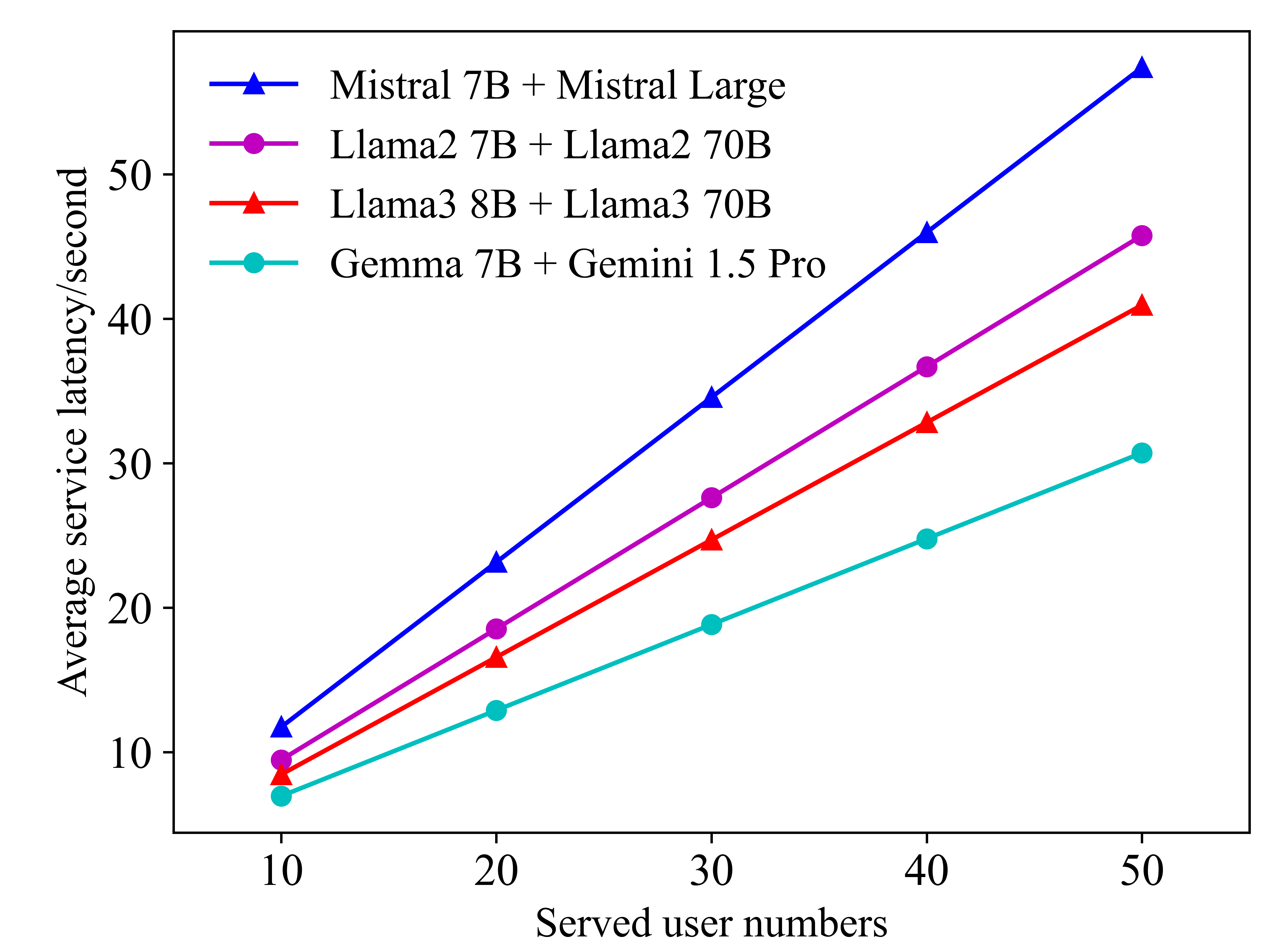} \label{f-r4}
}
\subfigure[{Average service delay with different prompt size using different LLM combinations.}]{
\includegraphics[width=5.2cm,height=3.7cm]{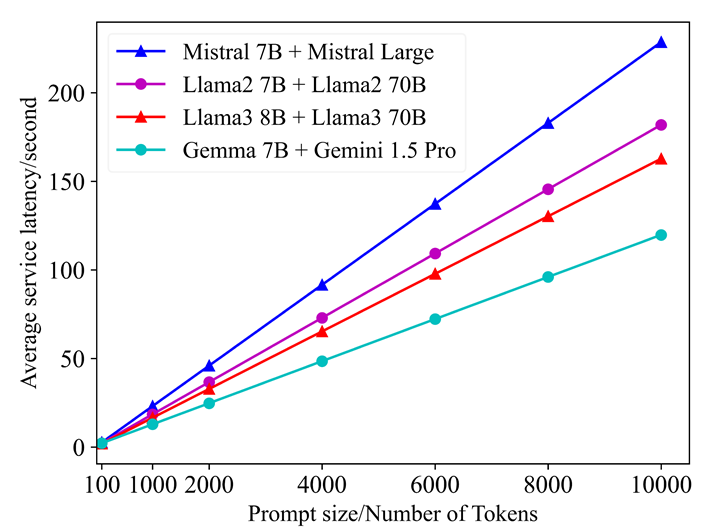} \label{f-r5}
}
\subfigure[{Average service delay with different tasks proportions using different LLM combinations.}]{
\includegraphics[width=5.2cm,height=3.7cm]{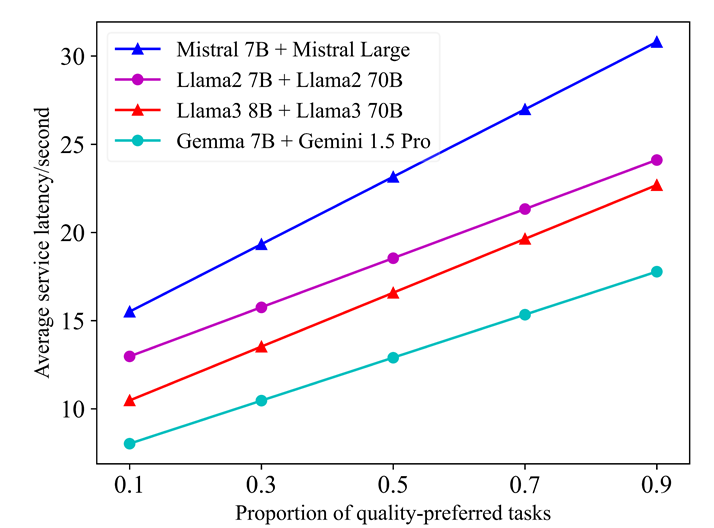} \label{f-r6}
}
\caption{Simulation results and comparisons}
\label{fig_all}
\vspace{-15pt}
\end{figure*}

\subsection{Simulation Settings}
We assume 20 users are randomly distributed in a cell based on 3GPP urban networks.
{The mobile users have two kinds of generation tasks: regular tasks such as chatting, translation, and summarization, and quality-preferred tasks such as scientific knowledge reasoning and coding. Such quality-based task classification aligns well with the evaluation schemes for LLMs, e.g., Chatbot Arena and Multi-task Language Understanding (MMLU).} 
The average output size is 1,000 tokens, and each token corresponds to about 4 bytes according to an OpenAI report \cite{token_size}.
We evaluate 2 LLMs for decision-making: Llama3-8B as a small-scale model that can be deployed at the network edge and GPT-4o as the latest large-scale LLM model as a benchmark for LLM-enabled methods.
For the network settings, we assume the edge LLM's TTFT and TPOT values are 0.23 and 1/75 second, while the values for the cloud LLM are 0.42 and 1/32 \cite{llm_website}.

\subsection{Simulation Results}

Fig. \ref{f-r1} shows the system reward of different tasks using Llama3-8B and GPT-4o, in which all tasks converge to a stable reward after exploration.
Here, the reward values indicate the overall service performance as defined in (\ref{eq12}). 
Llama3-8B and GPT-4o also achieve comparable performance as conventional DRL-based optimal baseline, demonstrating the potential of LLM-based optimization techniques.
In addition, Fig. \ref{f-r2} presents the service success rate for quality-preferred tasks, which means that the service score constraint in (\ref{e5:main}) should be fulfilled. It shows that the proposed in-context learning method can offload tasks properly to satisfy user requirements.

{We further compare the experience replay and exploration methods. 
in Fig. \ref{f-r3}. It shows that the proposed prioritized experience replay method can obtain comparable performance as optimal DRL baseline values.} 
{Specifically, epsilon-greedy exploration can try different decisions and collect good examples. These good examples will then be used in prioritized experience replay to guide the future decisions of LLMs. By contrast, latest experience reply and no-exploration methods present lower rewards as shown in the lower part of Fig. \ref{f-r3}.}

%

{Finally, Figs. \ref{f-r4}, \ref{f-r5} and \ref{f-r6} present the service delay of various LLMs under different user numbers, size of output tokens, and task distributions. 
In particular, when the proportion of quality-preferred tasks increases, these generation tasks must be offloaded to large-scale LLMs at cloud, leading to higher service delays. 
Meanwhile, Gemma 7B + Gemini 1.5 Pro can achieve the lowest overall delay than other combinations. This is because of the low TTFT values of these two LLMs, which are 1/155 second for Gemma 7B and 1/58 second for Gemini 1.5 Pro. Other LLMs have much higher values, i.e., 1/89 second for Llama2-7B and 1/40 second for Llama2-70B\cite{llm_website}. 
}
{In addition, some supplementary experiment results and analyses can be found in \cite{supple}. }


\section{Conclusion}
GAI is a promising technique for future 6G networks, and this work investigates the generation task offloading problems in 6G edge-cloud. 
It proposes a novel in-context learning method for generation task offloading, and the simulations demonstrated that the proposed technique can achieve satisfactory generation service quality.
{It is worth noting that such an edge-cloud deployment method may also increase the vulnerabilities of LLMs, e.g., inference and extraction attacks. In the future, we will explore secure and robust deployment strategies for LLMs in 6G networks, better guaranteeing the generation service reliability and quality.}

\normalem
\bibliographystyle{IEEEtran}
\bibliography{Reference}

\end{document}